# Clustering Approaches for Financial Data Analysis: a Survey

Fan Cai, Nhien-An Le-Khac, M-Tahar Kechadi,
*School of Computer Science & Informatics, University College Dublin, Ireland*

*Abstract*—Nowadays, financial data analysis is becoming increasingly important in the business market. As companies collect more and more data from daily operations, they expect to extract useful knowledge from existing collected data to help make reasonable decisions for new customer requests, e.g. user credit category, confidence of expected return, etc. Banking and financial institutes have applied different data mining techniques to enhance their business performance. Among these techniques, clustering has been considered as a significant method to capture the natural structure of data. However, there are not many studies on clustering approaches for financial data analysis. In this paper, we evaluate different clustering algorithms for analysing different financial datasets varied from time series to transactions. We also discuss the advantages and disadvantages of each method to enhance the understanding of inner structure of financial datasets as well as the capability of each clustering method in this context.

## I. INTRODUCTION

TODAY, we have a deluge of financial datasets. Faster and cheaper storage technology allows us to store ever-greater amounts of data. Due to the large sizes of the data sources it is not possible for a human analyst to come up with interesting information (or patterns) that will help in the decision making process. Global competitions, dynamic markets, and rapid development in the information and communication technologies are some of the major challenges in today's financial industry. For instance, financial institutions are in constant needs for more data analysis, which is becoming more very large and complex. As the amount of data available is constantly increasing, our ability to process it becomes more and more difficult. Efficient discovery of useful knowledge from these datasets is therefore becoming a challenge and a massive economic need.

On the other hand, data mining (DM) is the process of extracting useful, often previously unknown information, so-called knowledge, from large datasets (databases or data). This mined knowledge can be used for various applications such as market analysis, fraud detection, customer retention, etc. Recently, DM has proven to be very effective and profitable in analysing financial datasets [1]. However, mining financial data presents special challenges; complexity, external factors, confidentiality, heterogeneity, and size. The data miners' challenge is to find the trends quickly while they are valid, as well as to recognize the time when the trends are no longer effective. Moreover, designing an appropriate process for discovering valuable knowledge in financial data is a very complex task.

Different DM techniques have been proposed in the literature for data analysing in various financial applications. For instance, decision-tree [2] and first-order learning [3] are used in stock selection. Neural networks [4] and support vector machine [5] techniques were used to predict bankruptcy, nearest-neighbours classification [6] for the fraud detection. Users also have used these techniques for analysing financial time series [7], imputed financial data [8], outlier detection [9], etc. However, there are not many clustering techniques applied in this domain compared to other techniques such as classification and regression [2].

In this paper, we survey different clustering algorithms for analysing different financial datasets for a variety of applications; credit cards fraud detection, investment transactions, stock market, etc. We discuss the advantages and disadvantages of each method in relation to better understanding of inner structure of financial datasets as well as the capability of each clustering method in this context. In other words, the purpose of this research is to provide an overview of how basic clustering methods were applied on financial data analysis.

The rest of this paper is organised as follows. In Section II, we present briefly different financial data mining techniques that can be found in the literature. Section III describes briefly different clustering techniques used in this domain. We evaluate and discuss the advantages and disadvantages of these clustering methods in Section IV. We conclude and discuss some future directions in Section V.

## II. DATA MINING IN FINANCE

### A. Association Rules

Association Rule is a DM technique known as association analysis, which is useful for discovering interesting relationships hidden in large datasets. These relationships can be represented in the form of association rules or sets of frequent itemsets [2]. This technique can be applied to analyse data in different domains such as finance, earth science, bioinformatics, medical diagnosis, web mining, and scientific computation.

In finance, association analysis is used for instance in

N-A. Le-Khac: School of Computer Science & Informatics, University College Dublin, Ireland (**Corresponding author**: an.lekhac@lucd.ie).
F. Cai: School of Computer Science & Informatics, University College Dublin, Ireland (caifan.home@gmail.com).
M-T. Kechadi: School of Computer Science & Informatics, University College Dublin, Ireland (tahar.kechadi@ucd.ie).

customer profiling that builds profiles of different groups from the company's existing customer database. The information obtained from this process can help understanding business performance, making new marketing initiatives, analysing risks, and revising company customer policies. Moreover, loan payment prediction, customer credit policy analysis, marketing and customer care can also perform association analysis to identify important factors and eliminate irrelevant ones.

### B. Classification

Classification is another DM approach, which assigns objects to one of the predefined categories. It uses training examples, such as pairs of input and output targets, to find an appropriate target function also known informally as a classification model. The classification model is useful for both descriptive and predictive modelling [2]. In finance, classification approaches are also used in customer profiling by building predictive models where predicted values are categorical. Financial market risk, credit scoring/rating, portfolio management, and trading also apply this approach to group similar data together.

Classification can be considered as one of the important analytical methods in computational finance. Rule-based methods [2][3] can be used for the stock selection. Besides, bankruptcy prediction can use its geometric methods [4][5] where classification functions are represented with a set of decision boundaries constructed by optimising certain error criteria. Other methods such as Naïve Bayes classifiers [10], maximum entropy classifiers [11] were applied in bond rating and prototype-based classification methods such as nearest-neighbours classification was moreover used for the fraud detection.

### C. Clustering

Like classification, cluster analysis groups similar data objects into clusters [2], however, the classes or clusters were not defined in advance. Normally, clustering analysis is a useful starting point for other purposes such as data summarisation. A cluster of data objects can be considered as a form of data compression. Different domains can apply clustering techniques to analysis data such as biology, information retrieval, medicine, etc. In the business and finance, clustering can be used, for instance, to segment customers into a number of groups for additional analysis and marketing activities. As clustering is normally used in data summarisation or compression, there are not many financial applications that use this technique compared to classification and association analysis. We will survey some approaches in Section III.

### D. Other methods

Other mining techniques that can be applied for financial datasets are grouped in three categories: optimization, regression and simulation. For instance, portfolio selection, risk management and asset liability management can use different optimisation techniques such as genetic algorithms [12], dynamic programming [13], reinforcement learning [14], etc. Besides, linear regression [2] and wavelet regression [15] are popular methods in the domain of financial forecasting, option pricing and stock prediction.

### III. CLUSTERING METHODS

#### A. Partitioning Methods

K-means clustering [16] method aims to partition **n** observed examples into **k** clusters. Each example belongs to one cluster. All examples are treated with the equal importance and thus a mean is taken as the centroid of the observations in the cluster. With the predetermined **k**, the algorithm proceeds by alternating between two steps: assignment step and update step. Assignment step assigns each example to its closest cluster (centroid). Update step uses the result of assignment step to calculate the new means (centroids) of newly formed clusters. The convergence speed of the k-means algorithm is fast in practice but the optimal **k** value is not known in advance.

In [17], the author uses k-means algorithm to categorise mutual funds. The created clusters are assigned according to self-declared investment objectives and are compared to explain the difference between expectation and financial characteristics. Besides, in order to determine the number of clusters (k), the author applied the Hartigan's theory by evaluating the following formula:

$$\left( \frac{\sum_{i=1}^{k} ESS}{\sum_{i=1}^{k+1} ESS} - 1 \right) \times (n - k - 1) > 10 \qquad (1)$$

where k is the result with k clusters and ESS represents the sum of squares and n is the dataset's size. The number of clusters is the minimum *k* such that (1) is false.

#### B. Density-based

Another clustering approach is density based [2] which does not partition the sample space by mean centroid, but instead density based information is used, by which tangled, irregular contoured but well distributed dataset can be clustered correctly.

OPTICS [18] is a density based clustering technique to get insight into the density distribution of a dataset. It makes up for the weakness of the k-means algorithm for lack of knowledge of how to choose the value k. OPTICS provides a perspective to look into the size of density-based clusters.

Unlike centroid-based clustering, OPTICS does not produce a clustering of a dataset explicitly from the first step. It instead creates an augmented ordering of examples based on the density distribution. This cluster ordering can be used by a broad range of density-based clustering, such as DBSCAN. And besides, OPTICS can provide density information about the dataset graphically by cluster reachability-plot [18], which makes it possible for the user to understand the density-based structure of dataset.

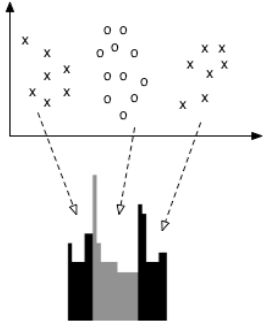

Figure I. 2-D dataset sample and corresponding reachability plot

Figure I gives the reachability-plot of the 2 dimensional dataset and the number of the valleys indicate that there are 3 density-based clusters.

However, OPTICS needs some priori, such as neighbourhood radius ($\varepsilon$) and a minimum number of objects (MinPts) within $\varepsilon$, by which directly density-reachable, density-connected, cluster and noise are defined as in [18].

DBSCAN [19] is based on density-connected range from arbitrary core objects, which contains MinPts objects in $\varepsilon$-neighbourhood. In OPTICS, cluster membership is not recorded from the start, but instead the order in which objects get clustered are stored. This information consists of two values: core-distance and reachability-distance. For more details on DBSCAN and OPTICS ordered dataset are provided in [18].

Core-distance of an object p is defined as:

$$core-distance_{\varepsilon,MinPts}(p) = \begin{cases} Undefined, if\ |neighbour_{\varepsilon}(p)| < MinPts \\ MinPts-distance(p), otherwise \end{cases}$$

Reachability-distance of an object q w.r.t object o is defined as:

$$reachability-distance_{\varepsilon,MinPts}(q,o)$$
$$= \begin{cases} Undefined, if\ |neighbor_{\varepsilon}(o)| < MinPts \\ \max(core-distance(o), distance(o,q)), otherwise \end{cases}$$

Since reachability plot is insensitive to the input parameters, [18] suggests that the values should be "large" enough to yield a good result with no undefined examples and reachability-plot looks not jagged. Experiments show that MinPts uses values between 10 and 20 always get good results with large enough $\varepsilon$. Briefly, reachability-plot is a very intuitive means to get the understanding of the density-based structure of financial data. Its general shape is independent of the parameters used.

### C. Data stream clustering

[9] applied an on-line evolving approach for detecting of financial statements' anomalies. The on-line evolving method [20] is a dynamic technique for clustering data stream. This method dynamically increases the number of clusters by calculating the distance between examples and existing cluster centres. If this distance is higher than a threshold value, a new cluster is created and initialized by the example. This clustering algorithm can be summarised in three main steps:

(1) Calculate the distance $D_{iJ}$ between data object $x_i$ to all existing cluster centres $Cc_J$, find the minimum distance $D_{ik}$ and compare it to the radius $R_k$ of cluster $C_k$.

(2) If $D_{ik} < R_k$ then $x_i$ belongs to cluster $C_k$, else find the nearest cluster $C_a$ and evaluate $S_a = D_{ia} + R_a$ against a threshold $\delta$.

(3) If $S_a > \delta$ then create a new cluster for $x_i$ else $x_i$ belongs to cluster $C_a$ and update $R_a = S_a/2$.

In this algorithm, the number of clusters is not predefined. However, the distance calculation and the threshold value needs expert to provide prior knowledge and so does label of newly formed cluster.

[21] applied a hierarchical agglomerative clustering [2] approach to analyse stock market data. The authors proposed an efficient metric for measuring the similarity between clusters; a key issue for hierarchical agglomerative clustering methods. This similarity between two clusters $C = \{C_1, C_2, ...C_k\}$ and $C' = \{C'_1, C'_2...C'_k\}$ is defined as follows:

$$Sim(C,C') = (\sum_i \max_j Sim(C_i,C'_j))/k$$

where

$$Sim(C_i,C'_j) = 2\frac{|C_i \cap C'_j|}{|C_i| + |C'_j|}$$

The authors also mentioned that some pre-processing techniques such as mapping, dimensionality reduction and normalisation should also be applied to improve the performance. Moreover, they used Precision-Recall method [21] to increase the cluster quality.

[7] also applied [21]'s approach for analysing financial data i.e. stock market. Besides, the authors defined a new distance metric based on the time period to cope with time series data. Concretely, the distance between stock i and stock j is given by:

$$d(i,j) = \|P_i - P_j\|_2$$

where

$$P_i(t) = \frac{s_i(t+1) - s_i(t)}{s_i(t)} \times 100$$

$s_i(t)$ is the stock value $i$ at time $t$. The authors stated that hierarchical agglomerative clustering fed by normalised percentage change after filtering outliers gives the best result. However the identification of outliers needs a priori threshold. Moreover, the authors combine neural networks and association analysis with the clustering technique to analyse stock market datasets.

## IV. EVALUATION AND ANALYSIS

### A. Datasets

Different financial datasets have been discussed in this section. Some of the $R_{i,j} = \frac{S_i + S_j}{M_{i,j}}$ were selected by the authors' approaches. For instance, [17] used data obtained from Morningstar including 904 different funds classified in seven different investment objectives: World Wide Bonds, Growth, SMEs, Municipal NY, Municipal CA, Municipal State and Municipal National. Each fund has 28 financial variables and all are normalised before analysis. Meanwhile, [9] used synthesis datasets with 1000 documents containing financial statements. In [21] the authors used Standard and Poor 500 index historical stock dataset. There are 500 stocks with daily price and each stock is a sequence of some length $l$ where $l \leq 252$. In [7] they analysed stock price datasets from 91 different stocks, which can be found at link http://finance.yahoo.com. The data covers three years; from November 1, 1999 to November 1, 2001.

We analyse moreover two financial datasets with k-means and density-based clustering approaches: German credit card and Churn. Both of these datasets are provided by UCI machine learning repository [22]. German credit dataset contains clients described by 7 numerical and 13 nominal attributes to good or bad credit risks. The data contains 1000 sample cases. The Churn dataset is artificial but are claimed to be similar to real-world measurement. It concerns telecommunications churn and contains 5 nominal attributes, 15 numerical attributes and 3333 examples. We analyse the dataset without the help of nominal attributes for several reasons, e.g. numerical attributes are taken internally within the commercial activities or business market while nominal attributes are stated by external concepts defined by market experts, whose significance is not promised. Moreover, nominal attributes are usually hierarchically dependent and can be missing while data mining models should have the capability to bypass these optional constraints to understand the structure of sample cases.

### B. Criteria

The criteria used to evaluate clustering methods depend on each approach. For instance, [17] applied a relevant value of k by using the formula (1) and then discuss on results obtained from the running of k-means algorithm to classify mutual funds.

[7] uses normalised change $P_i(t)$ of stock i to overcome the discrete essence of time and difficulties to treat deviations or first difference of prices due to the wide range of possible stock prices. External clustering statics such as entropy and purity are used to define the closeness within an industry, and internal statistics such as separation and the silhouette coefficient to tell what degree the industries' are separate from each other.

[9] does not give a clustering criterion but claims that their work is the first step to building robust financial statements' anomaly detection system but it highly depends on the operator monitoring the process.

In this paper, we use well-known internal criteria to evaluate the clustering behaviour. Davies-Bouldin Index (DBI) [23] is used as a first internal criterion for clustering, which is defined as follows:

$$DBI = \frac{1}{N} \sum_{i=1}^{N} D_i$$

where N is the number of clusters and $D_i$ is the tightness criteria of a cluster $C_i$, which takes the worst case scenario and is defined as:

$$D_i = \max_{j; i \neq j} R_{i,j}$$

where i and j are cluster indexes, $R_{i,j}$ is summary evaluation of two clusters of ratio between sum of tightness of two clusters and looseness between two centres.

$S_k$ is the average internal Euclidean distance of the cluster indexed by k, and $M_{i,j}$ is the Euclidean distance between two clusters.

$$S_i = \sqrt[2]{\frac{1}{T_i} \sum_{j=1}^{T_i} |X_j - A_i|^2}$$

$$M_{i,j} = \|A_i - A_j\|_2$$

where $A_i$ is the centroid of the cluster $C_i$, $T_i$ is the size of $C_i$, $X_i$ is an n dimensional feature vector assigned to $C_i$. The smaller DBI value is, the more efficient clustering is.

Dunn index (DI) is used as a second internal criteria for clustering, which is defined by:

$$DI = \min_{1 \leq i \leq N} \left\{ \min_{1 \leq j \leq N, j \neq i} \left\{ \frac{\delta(A_i, A_j)}{\max_{1 \leq k \leq N} \Delta_k} \right\} \right\}$$

where $\Delta_k$ is various types of size notation of a cluster, it could be farthest two points in side a cluster, mean distance between all pairs or distance of all the points from the mean.

$$\Delta_k = \max_{x,y \in C_k} |x - y|$$

and $\delta(A_i, A_j)$ is the closest distance between clusters

$$\delta(A_i, A_j) = \min_{x_i \in C_i, x_j \in C_j, i \neq j} |x_i - x_j|$$

Unlike DBI, the larger DI is the better is the clustering. It evaluates the inter-cluster and intra-cluster distances. However, like DBI, the best clustering loses most general structural information about the dataset.

The main difference between DBI and DI is that DBI indicates the average tightness while DI is a worst-case indicator.

## C. Partitioning Methods

As in [17] group mutual funds with different investment objectives, they claimed that cluster analysis is able to explain non-linear structural relationships among unknown structural dataset. They found that over 40% of the mutual funds do not belong to their stated categories, and despite the very large number of categories stated; three groups are very important. Clustering helps simplifying the financial data classification problem based on their characteristics rather than on labels, such as nominal labels (customer gender, living area, income or the success of the last transaction, etc.). Besides, nominal labels may be missing or not provided. Thus our effort is to understand the detailed structure of financial data classification without the given class labels.

We give the DBI and DI of K-Means clustering of both normalised and un-normalised two datasets (German credit dataset and Churn dataset) to figure out what are the optimal k values for given datasets. To avoid information overfitting and loss of generality, we test k from 2 to 20. We normalise the attributes values between [0:1] in order to avoid large-scale attributes dominating the dataset features.

$$x' = \frac{x - x_{min}}{x_{max} - x_{min}}$$

where the $x_{max}$ and $x_{min}$ are the max and min value of rescaled attribute.

From Figure II, k=12 is optimal by DBI and k=8 is the optimal value by DI for the original German credit dataset, k=8 is the optimal value for the normalised German credit dataset by both DBI and DI. From the result, we know that attribute scale affects the clustering evaluation since the DI of clustering original dataset is around 0. Normalisation unifies the results of both average tightness and worst case.

From Figure III, k=12 is optimal by DBI and k=17 by DI for original churn dataset. k=2 is the optimal value by both DBI and DI for normalised dataset. Again, we notice that normalisation unifies the optimal clustering scheme while original attribute scale giving two clustering solutions.

Figure V shows that normalised German credit dataset is well density distributed. When MinPts=10, by setting reachability-distance equal to 0.33, the dataset is partitioned into 23 density-based clusters and 1 noise cluster. There are 841 valid examples and 159 noise examples. When MinPts = 20, with the same reachability distance, dataset is partitioned into 15 density closed clusters and 1 noise cluster. There are 681 valid examples and 319 noise examples.

Despite the visualization of density distribution, from Table III, the clustering suffers from large proportion of noise and larger DBI values and lower DI values compared to K-means clustering. We can conclude that German credit dataset is more suitable for centroid-based clustering rather than density-based clustering.

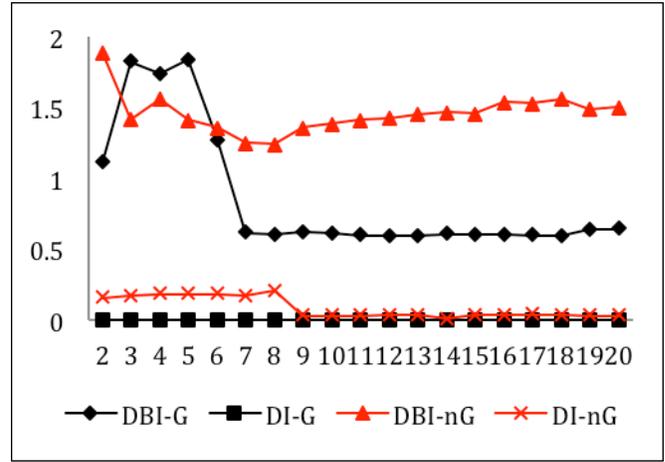

Figure II. DBI and DI of K-means clustering German dataset

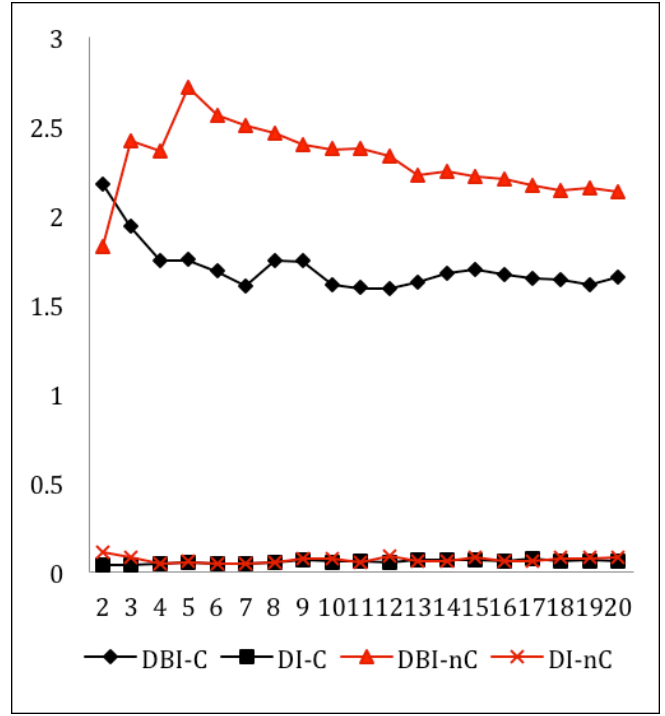

Figure III. DBI and DI of K-means clustering churn dataset

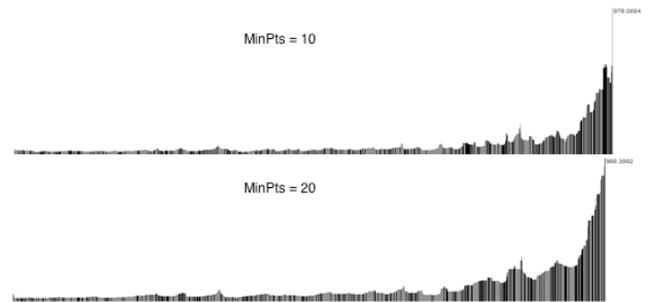

Figure IV. Reachability-plot of original German credit dataset

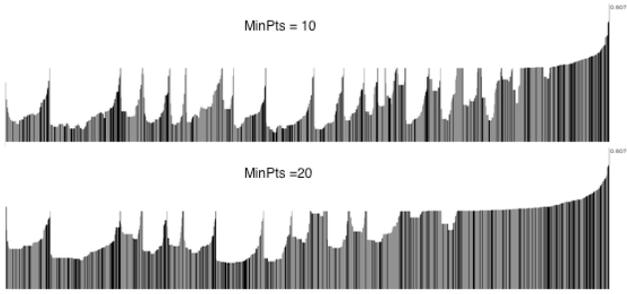

Figure V. Reachability-plot of normalized German dataset

Table III. DBSCAN clustering for normalized German credit dataset

| Reachability distance | MinPts | Noise | DBI | DI |
|---|---|---|---|---|
| 0.33 | 10 | No | 2.529 | 0.236 |
| | | Yes | 2.843 | 0.033 |
| 0.33 | 20 | No | 2.465 | 0.250 |
| | | Yes | 2.793 | 0.020 |

Figure VI shows that original churn dataset cannot partitioned into clusters based on density; the entire dataset behaves as a whole.

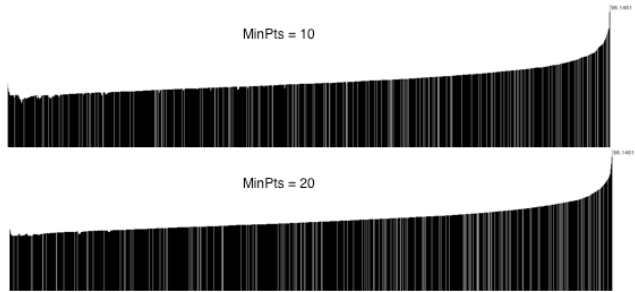

Figure VI. Reachability-plot of original Churn dataset

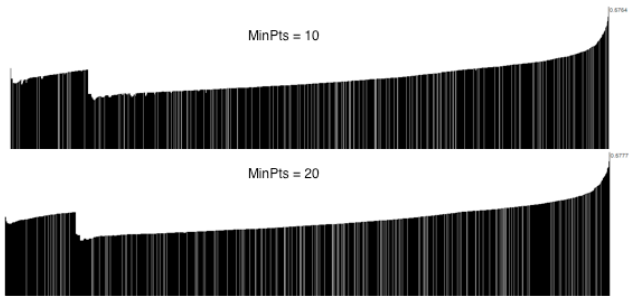

Figure VII. Reachability-plot of normalized Churn dataset

Figure VII shows that there are mainly two valleys when MinPts = 10 or 20, which indicates there are two incentive clusters in the churn dataset.

From Table IV, DBSCAN without noise examples gets good DBI while getting poor DBI with noise. However, DBSCAN clustering suffers from large proportion of noise again, which has over 980 noise examples (around 30% of noise). For financial dataset, noise should be very small and the data recorded should be generally trusted. Financial datasets are not usually density distributed, and therefore, density-based clustering is not appropriate.

Table IV. DBSCAN clustering and DBI for Churn dataset

| Reachability distance | MinPts | Noise | DBI | DI |
|---|---|---|---|---|
| 0.32 | 10 | No | 1.596 | 0.182 |
| | | Yes | 3.568 | 0.106 |
| 0.33 | 20 | No | 1.572 | 0.195 |
| | | Yes | 4.435 | 0.080 |

### D. Data stream clustering

In [9] the authors use an on-line evolving clustering to update the parameters: cluster number and cluster radius. Two levels of anomalies detection have different financial statement features. The first level is based on internal information related to the account, e.g. equipment, employee, etc. For every combination of the two parameters, at least one cluster is created. But the authors do not give a good reason for it. The second level is based on document type. However the distances among different types are different, which is a prior knowledge from expert as well. The threshold values for creating new clusters are determined by the experts for the first level and pre-defined distance for the second level. The monitoring process involves experts heavily to approve or disapprove the documents as well. The authors categorise their method as the first step to anomalies detection. They are committed to reduce the reliance on experts and combine off-line and on-line approaches in the future work.

In [7] the authors use hierarchical agglomerative clustering for the time-based normalised stock market data. Percentage change is chosen to be a good comparative measure and time-based normalisation is used to remove the overall trend of stock market and improve the accuracy caused by outliers. The approach removes all items as outliers if the average normalised distance across all the items exceeds a specified threshold, which requires domain expert knowledge. Moreover, the degree of correlation of time-series is decided in advance. The authors found complete link and Ward's Method performs reasonably well by better purity and filtering out fewer outlier stocks. By treating the outlier, the overall purity decrease only about 6%, the author claims time-series clustering can determine the industry classification given the historical price record of a stock.

However, we notice that data stream clustering needs too much prior or domain knowledge and a lot of tuning for different features of even a single domain. Clustering approaches of different fields are different in essence. Thus clustering is a good method to understand the financial time-series classification but not logically clear and efficient. Distance measure becomes even more complex due to time related nature because clustering does not have the capability to scale time related influence intelligently between examples. Experts have to determine that instead, e.g. length of periodicity, etc. Recurrent neural networks [24] and Gaussian Process [25] are more promising approaches and are more likely to handle time-series or periodical financial data classification.

## V. CONCLUSION AND FUTURE WORK

We show that density-based clustering does not suit financial dataset. Normalised centroid-based clustering with higher DI or lower DBI gives the best number of clusters to help understanding financial data classification. Original attribute scales do not reflect the behaviour similarity since Euclidean distance is dominated by large scaled attributes, best average tightness does not indicate the best case by departing the worst case. However, we still find some constrains, e.g., K-means clustering tends to find spherical clusters, centroid-based clustering does not handle the noise, etc.

This work can be seen as the first step to look into the structure of financial dataset by using clustering. We would further apply other techniques on financial datasets. This includes: (1) discover other centroid-based clustering approaches for financial datasets. (2) Find if nominal attributes are significant and introduce other criteria to evaluate the clusters. (3) Introduce weighted Euclidean distance instead of standard Euclidean distance to re-evaluate centroid-based clusters, as to overcome the limitations of K-means. (4) Introduce and compare different kinds of nonlinear classifiers to strengthen the recall and accuracy and improve prediction, interpretability of the results. These techniques include decision tree, nonlinear SVMs, different structures of neural networks and Gaussian processes with different kernel functions, etc.